\documentclass[12pt]{article}
\usepackage{amsmath,amssymb,amsthm,amscd}
\usepackage{epsf,amsfonts,hyperref}
\usepackage{graphicx}

\input epsf.sty
\newcommand{\CA}{{\cal A}}
\newcommand{\CB}{{\cal B}}

\newcommand{\CL}{{\cal L}}

\newcommand{\CN}{{\cal N}}

\def\IZ{{\mathbb Z}}

\newcommand{\re}{{\rm e}}
\newcommand{\ri}{{\rm i}}
\newcommand{\rd}{{\rm d}}

\newcommand{\be}{\begin{equation}}
\newcommand{\ee}{\end{equation}}
\newcommand{\ba}{\begin{aligned}}
\newcommand{\ea}{\end{aligned}}
\newcommand{\sectiono}[1]{\section{#1}\setcounter{equation}{0}}

\newcommand{\figref}[1]{Fig.~\protect\ref{#1}}
\renewcommand{\appendix}[1]{
    \addtocounter{section}{1}
    \setcounter{equation}{0}
    \renewcommand{\thesection}{\Alph{section}}
    \section*{Appendix \thesection\protect\indent #1}
    \addcontentsline{toc}{section}{Appendix \thesection\ \ \ #1}
}
\newcommand\encadremath[1]{\vbox{\hrule\hbox{\vrule\kern8pt
\vbox{\kern8pt \hbox{$\displaystyle #1$}\kern8pt}
\kern8pt\vrule}\hrule}}
\def\enca#1{\vbox{\hrule\hbox{
\vrule\kern8pt\vbox{\kern8pt \hbox{$\displaystyle #1$} \kern8pt}
\kern8pt\vrule}\hrule}}

\newcommand\figureframex[3]{
\begin{figure}[bth]
\hrule\hbox{\vrule\kern8pt \vbox{\kern8pt \vbox{
\begin{center}
{\mbox{\epsfxsize=#1.truecm\epsfbox{#2}}}
\end{center}
\caption{#3} }\kern8pt} \kern8pt\vrule}\hrule
\end{figure}
}
\newcommand\figureframey[3]{
\begin{figure}[bth]
\hrule\hbox{\vrule\kern8pt \vbox{\kern8pt \vbox{
\begin{center}
{\mbox{\epsfysize=#1.truecm\epsfbox{#2}}}
\end{center}
\caption{#3} }\kern8pt} \kern8pt\vrule}\hrule
\end{figure}
}

\newcommand{\beq}{\begin{equation}}
\newcommand{\eeq}{\end{equation}}
\newcommand{\bea}{\begin{eqnarray}}
\newcommand{\eea}{\end{eqnarray}}

\renewcommand{\and}{{\qquad {\rm and} \qquad}}

\newcommand{\virg}{{\qquad , \qquad}}

 \newcommand{\Tr}{{\,\rm Tr}\:}

\newcommand{\Res}{\mathop{\,\rm Res\,}}

\newcommand{\om}{\omega}

\newcommand{\Pint}{{\int\kern -1.em -\kern-.25em}}

\newcommand{\ovl}{\overline}

\newcommand{\acycle}{{\cal A}}
\newcommand{\bcycle}{{\cal B}}

\newcommand{\qbar}{\ovl{q}}

\newcommand{\bfa}{{\mathbf{a}}}

\newif\iffigs\figsfalse
\figstrue

\iffigs
  \input epsf
\else
\fi

\begin{document}

\begin{titlepage}
{}~
\hfill\vbox{
\hbox{CERN-PH-TH/2007-031}

\hbox{SPhT-T07/020}
}\break

\vskip .6cm

\centerline{\Large \bf
Holomorphic anomaly and matrix models}

\medskip

\vspace*{4.0ex}

\centerline{\large \rm Bertrand Eynard$^a$, Marcos Mari\~no$^b$\footnote{Also at Departamento de Matem\'atica, IST, Lisboa, Portugal}  and Nicolas Orantin$^a$}

\vspace*{4.0ex}

\centerline{ \rm ~$^a$ Service de Physique Th\'eorique de Saclay}
\centerline{F-91191 Gif-sur-Yvette Cedex, France}

\centerline{{\tt eynard@spht.saclay.cea.fr}, {\tt orantin@spht.saclay.cea.fr}}

\vspace*{1.8ex}

\centerline{ \rm ~$^b$Department of Physics, Theory Division, CERN}
\centerline{ \rm CH-1211 Geneva, Switzerland}

\centerline{\tt
marcos@mail.cern.ch}

\vspace*{6ex}

\centerline{\bf Abstract}
\medskip
The genus $g$ free energies of matrix models 
can be promoted to modular invariant, non-holomorphic amplitudes which only depend 
on the geometry of the classical spectral curve. We show that these non-holomorphic amplitudes satisfy the 
holomorphic anomaly equations of Bershadsky, Cecotti, Ooguri and Vafa. We derive as well 
holomorphic anomaly equations for the open string sector. These results provide 
evidence at all genera for the Dijkgraaf--Vafa conjecture relating matrix models to type B topological strings 
on certain local Calabi--Yau threefolds. 

\end{titlepage}
\vfill
\eject

\tableofcontents

\sectiono{Introduction and conclusions}

Topological string theory has been a fascinating laboratory to explore issues in string theory with important connections to other branches of physics and 
mathematics. The basic problem in topological string theory is to compute and understand closed and open string amplitudes on different geometric backgrounds. Particularly important 
among these are Calabi--Yau (CY) manifolds. Different techniques have been developed for the computation of these amplitudes. 
For type B topological strings on CY manifolds, a powerful method to solve the closed sector of the model are the holomorphic anomaly 
equations of \cite{bcov}. These equations control the $\bar t$-dependence of the closed string amplitudes $F^{(g)}(t, \bar t)$, and when combined with 
extra boundary conditions, they lead to explicit answers, see \cite{hkq,gkmw} for recent progress in this direction. 

On the other hand, in some special backgrounds one can use large $N$ dualities and geometric transitions to compute the holomorphic limit of $F^{(g)}(t, \bar t)$, which will be denoted by 
$F^{(g)}(t)$. In a groundbreaking paper \cite{dv}, Dijkgraaf and Vafa conjectured that on certain noncompact Calabi--Yau manifolds, 
where the geometry reduces to a complex curve, the $F^{(g)}(t)$ are given by the genus $g$ free energies of a matrix model in the $1/N$ expansion (see \cite{mmhouches} for 
a review). The curve encoding the CY geometry is then identified as the classical spectral curve of the matrix model. Given the connection between topological strings and type II superstrings, this has made possible 
to compute certain protected quantities in supersymmetric gauge theories by using matrix model technology. The connection between topological strings on noncompact 
CY manifolds and matrix models has been extended recently to the mirrors of toric geometries \cite{mm}.

The conjecture of Dijkgraaf and Vafa was verified at the planar level in \cite{dv}, and at genus one in \cite{kmt,dst}. In the simple case of the cubic matrix model, 
evidence for the conjecture was given at 
genus two in \cite{hk}, where it was shown that the matrix model expression for $F^{(g)}(t)$ is the holomorphic limit of a particular solution to the holomorphic 
 anomaly equations for the corresponding local curve. 
 
 In this paper we will give further 
evidence for the Dijkgraaf--Vafa conjecture and the related results of \cite{mm}, by using recent advances in the solution of matrix models at all orders in the $1/N$ expansion obtained in \cite{eynloopeq,EOloopeq, CEOloopeq} and more recently in \cite{eo}. One of the outcomes of 
these advances is that, as explained in \cite{eo}, one can extrapolate the matrix model procedure and define a series of amplitudes $F^{(g)}(t)$ and correlation functions 
$W_g(p_k)$ for any algebraic curve $H(x,y)=0$. When this curve is the classical spectral curve of a matrix model, one obtains in this way 
the $1/N$ expansion of the free energies and correlation functions. Using the results of \cite{eo}, we will show that 
it is possible to construct non-holomorphic free energies $F^{(g)}(t, \overline t)$ and correlation functions which satisfy the following conditions:

\begin{itemize}

\item In the holomorphic limit $\bar t \rightarrow \infty$ they reduce to the holomorphic amplitudes $F^{(g)}(t)$ associated to the spectral curve.

\item They are invariant with respect to the symplectic modular group of the curve. This is in contrast to their holomorphic limit, which does not have 
good modular properties \cite{eo}.

\item They satisfy the holomorphic anomaly equations of \cite{bcov}.

\end{itemize}

This gives a procedure to obtain the full 
non-holomorphic couplings $F^{(g)}(t, \bar t)$ in the context of matrix models, by using the requirement of modular invariance as in \cite{abk,dewit}, and shows that 
these couplings obey the equations of \cite{bcov} that characterize topological string amplitudes. This implies in particular that the matrix model free energy at genus $g$ (extended non-holomorphically in this way) must be equal to the type B free energy of the noncompact Calabi--Yau manifolds considered in \cite{dv,mm}, up to a holomorphic modular invariant quantity. 
Proving the Dijkgraaf--Vafa conjecture reduces now to proving that these holomorphic invariant functions vanish at all genera. 

An interesting spinoff of our work is that we can determine the holomorphic anomaly equations for {\it open} string amplitudes with respect to the closed moduli $\bar t$, 
at least in the local case. A direct string 
theory derivation of these equations has shown to be rather elusive \cite{bcov,ant}, and we expect them to be useful in future investigations of the open sector in the general 
case. The equations we find turn out to be closely related to the holomorphic anomaly equation for correlation functions obtained in \cite{bcov}, as we will explain in section 4. 

This work has other interesting implications. If one considers the hyperelliptic curves that appear in Seiberg--Witten theory as classical spectral curves of a matrix model,  
the amplitudes $F^{(g)}$ constructed with the recursive procedure of \cite{eo} should be closely related to the gravitational couplings of Seiberg--Witten theory 
introduced by Nekrasov in \cite{nekrasov}. It is known that these gravitational couplings can be promoted to non-holomorphic quantities that satisfy the holomorphic 
anomaly equations associated to the Seiberg--Witten curve \cite{hk,gkmw}. It follows from the results in this paper that these gravitational couplings differ from the 
$F^{(g)}$ of the spectral curve at most in a holomorphic, modular invariant object, and it is natural to conjecture that they are in fact equal. The existence of a matrix model description of Nekrasov's $F_g$ has been proposed in \cite{BertoldiHollowood}.

Our work also leaves some open problems. First of all, although it goes a long way towards proving the conjecture in \cite{dv}, it would be 
nice to finish the proof by fixing the remaining ambiguity. At a more conceptual level, the $\bar t$ dependence of the amplitudes has been introduced, 
following \cite{eo}, by requiring modular invariance and modifying 
the recursive construction of \cite{eo} accordingly. It would be very interesting to find 
a more intrinsic way to introduce this non-holomorphic dependence in the theories which are defined by matrix integrals. This might shed some light 
on the problem of background independence in topological string theory and of the wavefunction behavior of the topological string partition function \cite{witten}.

 The organization of this paper is as follows. In section 2 we review the holomorphic anomaly equations and the connections between topological strings and matrix models. Most of the 
 technology needed in this paper has been developed in \cite{eo}, and section 3 contains  some results of that paper which will be essential in our derivation. Section 4 contains the derivation 
 of the holomorphic anomaly equations. We actually provide a direct derivation and also a combinatorial derivation which makes contact with the recent results of \cite{abk}.

\sectiono{Holomorphic anomaly and topological strings}

\subsection{The holomorphic anomaly equations}

Topological string theory on Calabi--Yau threefolds (see \cite{mmhouches,nv,vonk}) for recent reviews) is defined by coupling 
the twisted $\CN=2$ sigma model to topological gravity and it comes in two versions: the A model 
(related to Gromov--Witten theory) and the B--model (related to deformations of complex structures). The topological 
amplitudes depend on a set of moduli that can be regarded as marginal deformations of the underlying $\CN=2$ theory. 
If we denote the chiral and antichiral operators of the theory that correspond to marginal directions
 by $\phi_I, \bar \phi_{\overline I}$, $I=1, \cdots, n$, we can deform 
the action in the standard way by using their descendants
\be
S_{\CN=2}=S(0) + \sum_{I=1}^n t^I \int_{\Sigma_g} \phi^{(2)}_I + \sum_{I=1}^n \overline t^{\overline I} \int_{\Sigma_g} \overline \phi^{(2)}_I
\ee
In the closed string sector we are interested in computing the genus $g$ free energy $F^{(g)}$ as well as the correlation functions 
of integrated chiral operators, 
\be
\label{corr}
C^{(g)}_{I_1 \cdots I_n}=\big\langle \int_{\Sigma_g} \phi^{(2)}_{I_1} \cdots \int_{\Sigma_g} \phi^{(2)}_{I_n}\big\rangle  
\ee
Since the $\overline t$ perturbation of the action can be written as a BRST-exact term with respect to the topological symmetry of the 
theory, one expects $F^{(g)}$ and the correlation functions to be $\overline t$-independent. However, as discovered in \cite{bcov1, bcov}, this is not the case 
once the twisted $\CN=2$ sigma model is coupled to gravity. The BRST-exact terms give a boundary contribution that can be evaluated 
recursively, and one finds holomorphic anomaly equations for the $\overline t$ dependence of the amplitudes. 

In order to write down the holomorphic anomaly equations, we need some basic ingredients from special geometry. The Zamolodchikov metric on the 
moduli space of marginal deformations is K\"ahler and can be derived from a K\"ahler function
\be
\label{zmetric}
G_{I \overline J} =\partial_I \partial_{\overline J} K. 
\ee
The amplitude at genus zero $F_0$, also 
called the prepotential, leads to a holomorphic three point function
\be
C_{IJK} =\frac{\partial^3 F_0}{\partial t^I \partial t^J \partial t^K}.
\ee
The holomorphic two--point coupling is given by 
\be
\label{taucoup}
\tau_{IJ}= \frac{\partial^2 F_0}{\partial t^I \partial t^J }
\ee
and plays an important role, as we will see later. Out of these quantities one can define a tensor
\be
C^{IJ}_{\overline K}=\re^{2K} G^{I \overline M} G^{J \overline N} {\overline C}_{\overline M \overline N \overline K}.
\ee
The Zamolodchikov metric (\ref{zmetric}) defines a connection on moduli space with Christoffel symbols
\be
\Gamma_{IJ}^K =G^{K \overline M}\partial_I G_{J \overline M}, \quad \Gamma_{\overline I \overline J}^{\overline K} =G^{M \overline K }\partial_{\overline I} G_{M  \overline J}.
\ee
It turns out that the $F^{(g)}$ are sections of a line bundle $\CL^{2-2g}$ on the moduli space, and the covariant derivative acting on them is given by 
\be
D_I =\partial_I -\Gamma_I +(2-2g) \partial_I K, 
\ee
where $K$ is the K\"ahler function appearing in (\ref{zmetric}). With these ingredients, the holomorphic anomaly equations of 
\cite{bcov} can be written as 
\be
\label{bcov}
\partial_{\overline K}  F^{(g)}  =\frac{1}{2}\,C^{IJ} _{\overline K} \Bigl(D_I D_J F_{g-1} + \sum_{h=1}^{g-1} D_I F_h D_J F_{g-h}\Bigr), \quad g\ge 2.
\ee
The equation for $g=1$ is slightly different and reads 
\be
\label{gone}
\partial_I \partial_{\bar J} F_1= \frac{1}{2} C_{IKL}C_{\overline J}^{KL} -\Bigl( {\chi \over 24}-1\Bigr) G_{I \overline J}.
\ee
Of course, the holomorphic anomaly equations determine $F^{(g)}(t, \overline t)$ only up to a holomorphic function of the moduli. This integration constant 
is called the holomorphic ambiguity. The holomorphic limit of $F^{(g)}(t, \overline t)$ is defined as
\be
\label{holimit}
F^{(g)}(t) =\lim_{\overline t \to \infty} F^{(g)}(t, \overline t). 
\ee
This is the limit which is for example appropriate to make contact with Gromov--Witten theory, in the A model. It has been observed \cite{dewit,abk} that the non-holomorphic 
couplings $F^{(g)}(t, \overline t)$ are invariant under the symmetry group acting naturally on the moduli space. In contrast, the holomorphic 
limit (\ref{holimit}) breaks the modularity properties.

We finally point out that the correlation functions (\ref{corr}) can be written as covariant derivatives of the $F^{(g)}(t, \overline t)$:
\be
\label{corrcov}
C^{(g)}_{I_1 \cdots I_n}=D_{I_1} \cdots D_{I_n} F^{(g)}, 
\ee
and they satisfy the holomorphic anomaly equation 
\be
\label{holocorr}
\ba
\partial_{\overline K} C^{(g)}_{I_1 \cdots I_n}&={1\over 2} C^{MN} _{\overline K} \Bigl(C^{(g-1)}_{MN I_1\cdots I_n} + \sum_{r=0}^g \sum_{s=0}^n {1\over s! (n-s)!} 
\sum_{\sigma \in S_n} C^{(r)}_{M I_{\sigma(1)} \cdots I_{\sigma(s)}} C^{(g-r)}_{N I_{\sigma(s+1)} \cdots I_{\sigma(n)}} \Bigr) \\
&-(2g-2+n-1) \sum_{s=1}^n G_{\overline K I_s} C^{(g)}_{I_1 \cdots I_{s-1} I_{s+1} \cdots I_n}.
\ea
\ee

\subsection{The local Calabi--Yau case and the Dijkgraaf--Vafa conjecture}

A particularly interesting class of topological strings are those defined on non--compact Calabi--Yau manifolds. These are commonly referred to 
as local Calabi--Yau manifolds. In the local B model the Calabi--Yau geometry is encoded in a Riemann surface $\Sigma_n$ of genus $n$ \cite{selfdual} (this should 
not be confused with the worldsheet of the string), and the resulting structure is significantly simpler than in the global case. The number of moduli is given by $n$, the genus of the curve, 
and they can be parametrized by 
\be
\label{tint}
t^I \varpropto \int_{A^I} y(x) \rd x, \quad I=1, \cdots, n, 
\ee
where $A^I$ are the compact A-cycles of $\Sigma_n$, and $y(x)\rd x$ is a meromorphic one-form on $\Sigma_n$ which can be constructed from the equation for the 
spectral curve. The proportionality constant in (\ref{tint}) will be determined below by a detailed comparison with the matrix model 
results. The imaginary part of the two-point coupling (\ref{taucoup}) is now a positive 
definite matrix, and the Zamolodchikov metric is simply given by
\be
G_{I\bar J}= -\ri(\tau -\bar \tau)_{IJ}.
\ee
As noticed in \cite{kz,hk,abk}, the holomorphic anomaly equations also simplify in the local case. First of all, 
the tensor $C_{\overline K}^{IJ} $ entering in (\ref{bcov}) is now 
simply given by
\be
\label{tensorlocal}
C_{\overline K}^{IJ} =G^{I\overline M} G^{J \overline N} C_{\overline M \overline N \overline K}=- [(\tau-\bar \tau)^{-1}]^{IM} [(\tau-\bar \tau)^{-1}]^{JN} C_{\overline M \overline N \overline K}.
\ee
The covariant derivative acting on the $F^{(g)}$ amplitudes involves now only the Christoffel symbol (i.e. the term involving $\partial_I K$ is no longer present). Notice that the 
Christoffel symbol can be written as
\be
\label{chris}
\Gamma^L_{IJ} =[(\tau-\bar \tau)^{-1}]^{LM} \partial_I \tau_{J M},
\ee
With these modifications, the holomorphic anomaly for $F^{(g)}$, $g\ge 2$ is still given by (\ref{bcov}). For $g=1$ the last term of (\ref{gone}) is not present. The holomorphic 
anomaly equations for the correlation functions are given by (\ref{holocorr}) but the contribution of the last line is absent in the local case. 

Since it will be useful later one, we now write the holomorphic anomaly equations for $g\ge 2$, in the local case, in terms of a generating functional. We introduce, 
\be
Z=\exp\Bigl[ \sum_{g=0}^{\infty} N^{2-2g} F^{(g)}(t)\Bigr], 
\ee
where $N$ plays the role of a genus--counting parameter. We also define, 
\be
\tilde Z= \exp \bigl\{-N^2 F^{(0)} -F^{(1)}\bigr\} Z, \quad \hat Z= \exp \bigl\{-N^2 F^{(0)} \bigr\} Z.
\ee
For $g \ge 2$ we want to express the antiholomorphic derivative of $\tilde Z$ in terms of $\hat Z$. Using that 
\be
\partial^2_{IJ} \log \, \hat Z = {1\over \hat Z} \partial^2_{IJ} \hat Z -{1\over Z^2} \partial_I \hat Z \partial_J \hat Z
\ee
we can write (\ref{bcov}) in the local case as
\be\label{bcovbis}
{1 \over \tilde Z} \partial_{\overline K} \tilde Z =-{1 \over 2 N^2} {1\over \hat Z } C^{IJ} _{\overline K} \Bigl( \partial_{IJ}^2 \hat Z -\Gamma_{IJ}^K \partial_K \hat Z\Bigr).
\ee

In the local case, when the Calabi--Yau geometry reduces to a Riemann surface $\Sigma_n$ of genus $n$, 
the natural group acting on the moduli space is the mapping class group, and there is an induced action of the symplectic modular group 
\be
{\rm Sp}(2n, \IZ)
\ee
on the topological string amplitudes $F^{(g)}(t, \overline t)$ and their holomorphic limit. The non- holomorphic 
amplitudes $F^{(g)}(t, \overline t)$ of the local 
geometry turn out to be modular invariant under the action of this group \cite{hk,abk,dewit}, while the $F^{(g)}(t)$ do not have 
good modular properties.

An interesting class of local Calabi--Yau backgrounds are the ones described by the equation
\be
\label{cydv}
uv=H(x, y), \quad H(x,y)=y^2-(W'(x))^2+f(x) 
\ee
where $W(x), f(x)$ are polynomials of degree $d+1$, $d-1$ respectively. The Riemann surface $H(x,y)=0$ associated to this geometry is 
the hyperelliptic curve 
\be
\label{hypercurve}
y^2= (W'(x))^2-f(x)
\ee
of genus $n=d-1$. Dijkgraaf and Vafa conjectured in \cite{dv} (see \cite{mmhouches} for a review) that the holomorphic amplitudes 
$F^{(g)}(t)$ of type B topological string theory on the backgrounds of the form (\ref{cydv}) 
are given by the $1/N$ expansion of a matrix model with spectral curve (\ref{hypercurve}). This is 
in fact a matrix model with potential $W(x)$ and generically with $d-1$ cuts around its extrema.
This conjecture has been verified at the planar level in \cite{dv}, at genus one in \cite{kmt,dst}, and at genus two in \cite{hk}.

Another interesting class of local geometries is given by the mirrors of toric Calabi--Yau manifolds. In this case, 
the Riemann surface is a complex curve in ``exponentiated" variables which can be written in the form 
\be
\label{my}
y^2=M^2(x) \prod_{i=1}^{\ell} (x-x_i),
\ee
where $M(x)$ is in this case a transcendental function. As shown in \cite{eo} and reviewed below, one can generalize 
the matrix model idea and associate a series of free energies $F^{(g)}(t)$ and correlation functions to any algebraic curve. It was recently proposed in \cite{mm} that holomorphic 
open and closed type B topological string amplitudes for the mirrors of toric background are given by the correlation functions and free energies, respectively, associated to the spectral curve 
(\ref{my}) with the procedure of \cite{eo}. 

We then see that, for local CY backgrounds, where the geometry reduces to a complex curve, the holomorphic limit of topological string amplitudes in the B model has been conjectured 
to be given by the free energies and correlation functions associated to the complex curve by the recursive procedure of \cite{eo}. In the example (\ref{cydv}) this coincides with the $1/N$ expansion of the free energy and correlators of a matrix model with potential $W(x)$. In the rest of this paper we will give strong support to this conjecture by showing that the quantities defined in \cite{eo} can be extended to non-holomorphic objects which have good modular properties and satisfy the holomorphic anomaly equation (\ref{bcov}) in the local case.

\sectiono{Review of matrix models}

\subsection{Formal matrix models and algebraic geometry}

Random matrix models became very popular in the 80's and 90's after it was discovered by 't Hooft \cite{thooft} and then in \cite{BIPZ} that the Feynman graphs of 
Hermitian matrix integrals are in fact discrete surfaces.
Formal random matrix models are thus generating functions which count discrete maps of given topology.
For instance the 1 matrix integral:
\be\label{defz1MM}
Z_{\rm 1MM} = \int dM\,\, e^{- {N \over t} \Tr V(M) }
\ee
 counts discrete maps made of triangles, squares, ... $k$-gones where $k\leq \deg V$.
 The 2 matrix integral
\be\label{defz2MM}
Z_{\rm 2MM} = \int dM_1\, dM_2\,\, e^{- {N \over t} \Tr (V_1(M_1)+V_2(M_2) - M_1 M_2) }
\ee
 counts discrete maps made of polygons of two possible colors (or call it two possible spins $\pm$).
 It was introduced by Kazakov \cite{KazIsing} as an Ising model on a random map.
 
 In all cases, the formal integral is defined as a formal series in its small $t$ expansion, and it turns out that to any order in $t$, one has a ``topological expansion":
 \be
 \label{topexpansion}
F_{\rm MM}= \ln{Z_{\rm MM}} = \sum_{g=0}^\infty N^{2-2g} F^{(g)}
 \ee
 where $F^{(g)}$ is the generating function which counts maps of genus $g$.
(see \cite{eynform} for a review on this subject). 
 
\smallskip
 
The goal was then to compute explicitly the coefficients $F^{(g)}$ in that expansion. Many methods have been invented, but the only one which has really succeeded so far is the so-called loop equation method which just amounts to integrate by parts or equivalently write Ward identities or Virasoro (or W-algebra) constraints.
It was soon realized that loop equations allow in principle to find $F^{(g)}$ for any $g$ \cite{ACKM, Akemann}, but a more explicit solution (which can be represented diagrammatically) was found only recently first for the 1-matrix model \cite{eynloopeq}, then for the 2-matrix model \cite{EOloopeq, CEOloopeq}, and more 
recently for the matrix model with external field \cite{eo}.

Then it was understood in \cite{eo}, that the solution of matrix models loop equation extends beyond the context of matrix models and is in fact a property of algebraic geometry in general. The diagrammatic method of \cite{eo} allows to compute explicitly all the free energies $F^{(g)}$ appearing in (\ref{topexpansion}) 
%(generating functions of discrete closed surfaces of genus $g$):
%
 %\be
%F_{\rm MM}=\ln{Z_{\rm MM}} = \sum_{g=0}^\infty N^{2-2g} F^{(g)}
% \ee
as well as all correlation functions (generating functions of discrete open surfaces of genus $g$ with $k$ boundaries):
\be
 \left< \Tr {dx_1 \over x_1-M} \dots \Tr {dx_k \over x_k-M} \right>
= \sum_{g=0}^\infty   N^{2-2g-k}\, W_k^{(g)}(x_1, \dots x_k)
\ee
where the mean value refers to the measure in (\ref{defz1MM}) and the variables $x_i$ are fugacities associated to the lengths of boundaries.

\medskip

%{\bf summary of the construction of \cite{eo}:}

Let us now give a brief summary of the construction of \cite{eo}. 

Given an arbitrary algebraic curve $H(x,y)=0$, of genus $n$, one constructs recursively a sequence of multilinear symmetric meromorphic forms $W_k^{(g)}$ and some scalars $F^{(g)}$'s in terms of residues localized at branchpoints of the curve.

Many properties of those $W_k^{(g)}$'s and $F^{(g)}$'s are described in \cite{eo}, in particular variations of conformal structure, integrable structure, invariance under symplectic transformations of the curve, homogeneity, and the one which interests us now is modular transformations.

Indeed, the construction of \cite{eo} is based on the Bergmann kernel,
which is defined for a given canonical basis of cycles. When one changes the basis of cycles, the Bergmann kernel changes, and the $W_k^{(g)}$'s and $F^{(g)}$'s change accordingly.
In \cite{eo}, a "modified" Bergmann kernel was introduced in order to easily take into account those modular changes. The modified Bergmann kernel depends on an arbitrary complex symmetric matrix $\kappa$, and a modular change of cycles merely amounts (as far as only the Bergmann kernel is concerned) to a change of $\kappa$.
Since the $W_k^{(g)}$'s and $F^{(g)}$'s modular dependence is only in the Bergmann kernel, their modular properties are entirely encoded in their $\kappa$ dependence.
Thus in \cite{eo}, the following quantities were computed:
\be
{\partial W_k^{(g)} \over \partial \kappa} \qquad  {\rm and} \quad {\partial F^{(g)} \over \partial \kappa} 
\ee
and they showed a striking similarity with holomorphic anomaly equations.

The goal of this article is to prove that indeed one can derive the holomorphic anomaly equation from the construction of \cite{eo}.

\subsection{Variations of the matrix model free energies}

Let us consider an algebraic curve $H(x,y)$ of genus $n$ and a canonical basis of cycles on it:
\beq
\forall i,j = 1 \dots n
\virg
\underline\acycle_i\cap \underline\bcycle_j=\delta_{ij}
\virg
\underline\acycle_i\cap \underline\acycle_j=0
\virg
\underline\bcycle_i\cap \underline\bcycle_j=0.
\eeq
There are $n$ linearly independant holomorphic forms $\rd u_i$ on $H(x,y)$ normalized on the $\underline\acycle$-cycles:
\be
\forall i,j=1 \dots n \virg \oint_{\underline\acycle_i} \rd u_j = \delta_{ij}
\ee
and the Riemann matrix of period $\tau$ is a symmetric $n \times n$ matrix defined by
\be
\oint_{\underline\bcycle_j} \rd u_i = \tau_{ij}.
\eeq

\bigskip

{\bf\noindent  Bergmann kernel}

\smallskip

There exists a unique bilinear form $\underline{B}(p,q)$ with a unique double pole
at $p=q$ without residue and normalized on the $\underline{\acycle}$ cycles:
\beq
\underline{B}(p,q) \mathop{\sim}_{p \to q} {\rd z(p) \rd z(q) \over (z(p)-z(q))^2} \;  + \; \hbox{finite}
\virg
\oint_{\underline{\CA}} \underline{B} = 0.
\eeq
where $z$ is any local parameter. $\underline{B}$ is often referred to as the {\bf Bergmann kernel} in the literature.
For instance, on a torus, $\underline{B}$ is the Weierstrass function.

Let us now introduce a new set of cycles, depending on an arbitrary complex symmetric matrix $\kappa$:
\beq
\bcycle := \underline\CB - \tau \underline\CA
\virg
\CA := \underline\CA - \kappa \CB
\eeq
and  define a $\kappa-$modified Bergmann kernel normalized on these new
 cycles by the constraints:
\beq
B(p,q) \sim_{p \to q} {\rd z(p) \rd z(q) \over (z(p)-z(q))^2} \;  + \; \hbox{finite}
\virg
\oint_\CA B = 0.
\eeq
This definition implies the relation:
\beq
B(p,q) = \underline{B}(p,q) + 2i \pi\sum_{i,j} \rd u_i(p)\, \kappa_{ij}\, \rd u_j(q).
\eeq
Under a modular transformation:
\beq
\tau \to (C-\tau D)^{-1}\,(\tau E -F) \qquad , \,\,\, E^t C - F^t D = {\rm Id} 
\quad , \,\, C^t D = D^t C
\quad , \,\, E^t F = F^t E
\eeq
the Bergmann kernel $\underline{B}$ changes, as well as the $\kappa$-modified Bergmann kernel, and the change is equivalent to a change of $\kappa$:
\beq
\kappa \to  D (C-\tau D)^{-1}\,+ \left((C-\tau D)^{-1}\right)^t \kappa (C-\tau D)^{-1}
\eeq
Therefore, the action of a modular transformation is equivalent to a change of $\kappa$, and all modular properties of the Bergmann kernel, and thus of the $F^{(g)}$'s are encoded in the $\kappa$ dependence.

One can verify that if $\kappa=(\overline\tau-\tau)^{-1}$, then the Bergmann kernel is modular invariant (see for instance a proof in \cite{eo}).

\bigskip

{\bf\noindent  Filling fractions}

\smallskip

We also introduce the filling fractions
\beq
\epsilon_i := {1 \over 2 \ri \pi} \oint_{\underline\acycle_i} y \rd x .
\eeq

\bigskip

{\bf\noindent  Branch points}

\smallskip

Let us also consider the branch points $a_i$ defined by $\rd x(a_i)=0$. When a point $p$ approaches a branch point $a_i$, there
exists a unique point $\overline{p}$ such that $x(p)= x(\overline{p})$ and $\overline{p} \to a_i$.
Notice that $\overline{p}$ is defined locally near any branchpoint, but not globally (except for hyperelliptical curves of the form $y^2=P(x)$).

\bigskip

The free energies and correlation functions are recursively defined as follows:

\medskip
\noindent {\bf Correlation functions}
\beq
W_k^{(g)}=0 \quad {\rm if}\,\, g<0
\eeq
\beq
W_1^{(0)}(p) = 0
\eeq
\beq
W_2^{(0)}(p_1,p_2) = B(p_1,p_2)
\eeq
and
\be\label{defWkgrecursive}
\ba
& W_{k+1}^{(g)}(p,p_1,\dots,p_k) \\
&=\sum_i \Res_{q\to a_i} {\rd E_{q}(p)\over \om(q)}\,\Bigl(\sum_{m=0}^g \sum_{J\subset K} W_{j+1}^{(m)}(q,p_J)W_{k-j+1}^{(g-m)}(\qbar,p_{K/J})
+ W_{k+2}^{(g-1)}(q,\qbar,p_K) \Bigr)
\ea
\ee
with the notations ($q$ being near a branchpoint):
\be
\om(q) = (y(q)-y(\qbar))\rd x(q), \qquad 
\rd E_{q}(p) = {1\over 2} \int_{q}^{\qbar} B(\xi,p)
\eeq
where the integration path lies entirely in a vicinity of $a_i$. 
If $J=\{ i_1,i_2,\dots,i_j\}$ is a set of indices, we write $p_J=\{ p_{i_1},p_{i_2},\dots,p_{i_j}\}$, and in the equation above we have $K=\{1,2,\dots,k\}$ and the summation over $J$ is among all subsets of $K$.

It is important to notice that by construction, all $W_k^{(g)}$'s have vanishing $\acycle$-cycle integrals:
\beq
\oint_{\acycle} W_k^{(g)}=0
\eeq
which implies:
\beq\label{vanishacycle}
\oint_{\underline\acycle} W_k^{(g)} = \kappa \oint_{\bcycle} W_k^{(g)}
\eeq

\medskip
\noindent{\bf Free energies.}

For $g>1$
\beq
F^{(g)} = {1\over 2g-2}\,\sum_i \Res_{q\to a_i} \Phi(q) W_{1}^{(g)}(q).
\eeq
where $\Phi(q) = \int^q y \rd x$ is any antiderivative of $y \rd x$, i.e.
\beq
\rd \Phi = y \rd x
\eeq

{\bf remark:} it is proved in \cite{eo} that
\beq
W_k^{(g)} = {1\over 2g+k-2}\,\sum_i \Res_{a_i} \Phi W_{k+1}^{(g)}.
\eeq
so that $F^{(g)}$ can be regarded as $F^{(g)}= W^{(g)}_0$.

\bigskip
{\bf \noindent Modular transformations and $\kappa$ dependence.}
\smallskip

From their definitions, it is clear that the only modular dependence, and also the only $\kappa$ dependence of the $W_k^{(g)}$'s is through the Bergmann kernel, and therefore, a modular transformation is equivalent to a change of $\kappa$.
Moreover, each $W_k^{(g)}$ is a polynomial in $\kappa$ of degree at most $3g-3+2k$.
In order to study modular transformations of the $W_k^{(g)}$'s, or their $\kappa$ dependence, it is convenient to compute $\partial W_k^{(g)}/\partial \kappa$. This was done in \cite{eo} and the result is:
\be
\label{variatW}
\ba
2\ri\pi {\partial \over \partial \kappa_{ij}}\,W^{(g)}_{k}(p_K)
&=  {1\over 2}\,\oint_{r\in\bcycle_j}\oint_{s\in\bcycle_i} W^{(g-1)}_{k+2}(p_K,r,s) \\
& + {1\over 2}\,\sum_h \sum_{L\subset K} \oint_{r\in\bcycle_i} W^{(h)}_{|L|+1}(p_L,r) \oint_{s\in\bcycle_j} W^{(g-h)}_{k-|L|+1}(p_{K/L},s) 
\ea
\ee
and in particular for $k=0$:
\be\label{variatF}
 2\ri\pi {\partial \over \partial \kappa_{ij}}\,F^{(g)}
=  {1\over 2}\,\oint_{r\in\bcycle_j}\oint_{s\in\bcycle_i} W^{(g-1)}_{2}(r,s) + {1\over 2}\,\sum_{h=1}^{g-1} \oint_{r\in\bcycle_i} W^{(h)}_{1}(r) \oint_{s\in\bcycle_j} W^{(g-h)}_{1}(s)
\qquad g\geq 2.
\ee

Also, notice that if $\kappa=(\overline\tau-\tau)^{-1}$, then all $W^{(g)}_{k}$'s and $F^{(g)}$'s (with $2g+k\geq 2$) are modular invariant.

\subsubsection{Variations wrt filling fractions}

The variation of the Bergmann kernel wrt filling fractions follows from Rauch variational formula, and it was computed in \cite{eo} that (in fact in \cite{eo}, $\kappa$ was considered a constant, and thus we just have to add an extra ${\partial \kappa/ \partial \epsilon}$ term):
\beq
{\partial B(p,q)\over \partial \epsilon} 
= \Res_{r\to \bfa} {B(r,p) B(r,q) \rd u(r)\over \rd x(r) \rd y(r)} - 2\ri\pi\, \rd u^t(p)  \kappa {\partial \tau\over \partial \epsilon}  \kappa  \rd u(q) + 2\ri\pi\, \rd u^t(p) {\partial \kappa\over \partial \epsilon}\rd u(q)
\eeq
which was written using an operator in \cite{eo}:
\beq
\Delta_{\epsilon} = \partial_\epsilon +  \Big(\kappa  {\partial \tau\over \partial \epsilon}  \kappa - {\partial \kappa\over \partial \epsilon}\Big) {\partial \over \partial \kappa}
\eeq
so that:
\beq\label{defDhat}
\Delta_\epsilon  B(p,q) 
= \Res_{r\to \bfa} {B(r,p) B(r,q) \rd u(r)\over \rd x(r) \rd y(r)} 
\eeq
This property was sufficient to ensure in \cite{eo}, that:
\beq\label{DhatWkgeo}
\Delta_\epsilon W_k^{(g)} = - \oint_{\bcycle}  W_{k+1}^{(g)}
\eeq

\sectiono{Holomorphic anomaly equations and matrix models}

In this section we give two proofs of the holomorphic anomaly equation for matrix models. 

\medskip

In the following, we consider the choice 
\be\label{choicekappa}
\kappa = - {1 \over \tau - \overline{\tau}}
\ee
in which case the $F^{(g)}$'s and $W_k^{(g)}$'s are modular invariant \cite{eo} but not holomorphic.

\subsection{A direct proof}

%\subsubsection{Free energy}

%Roughly speaking the free energies depend only on the choice of a Bergmann kernel. Thus one only needs to know the variations of the 
%Bergmann kernel to know the variations of the free energies.

We first consider the variation with respect to $\overline{\epsilon}$.
The only antiholomorphic dependence of $W_k^{(g)}$ on $\overline{\epsilon}$ comes from its
dependence on $\kappa=1/(\overline\tau-\tau)$, therefore
\beq
{\partial W_k^{(g)} \over \partial \overline{\epsilon}} = {\partial \kappa \over \partial \overline{\epsilon}}
{\partial W_k^{(g)} \over \partial \kappa}
\quad , \quad
{\partial \kappa \over \partial \overline{\epsilon}} = -\kappa {\partial \overline\tau\over \partial\overline\epsilon} \kappa 
\quad , \quad
{\partial^2 F^{(0)}\over \partial \epsilon^2} = 2\ri \pi \tau
\eeq
so that
\beq
{\partial W_k^{(g)} \over \partial \overline{\epsilon}} = 
-{1\over 2\ri \pi}\, \kappa {\partial^3 {\overline{F}}^{(0)} \over \partial {\overline\epsilon}^3} \kappa 
{\partial W_k^{(g)} \over \partial \kappa}.
\eeq
Then, using eq. (\ref{variatW}):
\bea
{\partial W_k^{(g)} \over \partial \overline{\epsilon}} &=& 
-{1\over (2\ri \pi)^2}\, \kappa {\partial^3 {\overline{F}}^{(0)} \over \partial {\overline\epsilon}^3} \kappa \,\, 
 {1\over 2}\,\oint_{r\in\bcycle_j}\oint_{s\in\bcycle_i} \Big( W^{(g-1)}_{k+2}(p_K,r,s) \cr
& & + \,\sum_h \sum_{L\subset K}  W^{(h)}_{|L|+1}(p_L,r) W^{(g-h)}_{k-|L|+1}(p_{K/L},s) \Big)
\eea

Notice that with $\kappa=1/(\overline\tau-\tau)$, which satisfies:
\beq
\kappa {\partial \tau\over \partial \epsilon}  \kappa  = {\partial \kappa\over \partial \epsilon}
\eeq
the differential operator $\Delta$ of eq. (\ref{defDhat}) reduces to the usual derivative $\Delta_{\epsilon_I} = \partial/\partial \epsilon_I$ , and therefore using eq. (\ref{DhatWkgeo}):
\beq
{\partial W_k^{(g)}\over \partial\epsilon} = - \oint_{\bcycle}  W_{k+1}^{(g)} = - \oint_{\underline\bcycle}  W_{k+1}^{(g)} +\tau \oint_{\underline\acycle}  W_{k+1}^{(g)}
\eeq
Acting again with $\partial_\epsilon$ we find (we use \ref{vanishacycle}):
\bea
\label{seconder}
{\partial^2 W_k^{(g)}\over \partial\epsilon^2} 
&=& \oint_{\bcycle} \oint_{\bcycle}  W_{k+2}^{(g)}   + {\partial \tau\over \partial\epsilon} \oint_{\underline\acycle}  W_{k+1}^{(g)} \cr
&=& \oint_{\bcycle} \oint_{\bcycle}  W_{k+2}^{(g)}   + {\partial \tau\over \partial\epsilon} \kappa \oint_{\bcycle}  W_{k+1}^{(g)} 
\eea
and therefore:
\beq
{\partial^2 W_k^{(g)}\over \partial\epsilon^2} 
+ {\partial \tau\over \partial\epsilon} \kappa {\partial W_k^{(g)}\over \partial\epsilon} = \oint_{\bcycle} \oint_{\bcycle}  W_{k+2}^{(g)}   
\eeq

Finally we get the equation:
\bea
{\partial W_k^{(g)} \over \partial \overline{\epsilon}} &=& 
-{1\over (2 \ri\pi)^2}\, \kappa {\partial^3 {\overline{F}}^{(0)} \over \partial {\overline\epsilon}^3} \kappa \,\, 
 {1\over 2}\,\Big( {\partial^2 W_k^{(g-1)}\over \partial\epsilon^2} + {\partial \tau\over \partial\epsilon} \kappa {\partial W_k^{(g-1)}\over \partial\epsilon}  \cr
& & + \,\sum_h \sum_{L\subset K}  {\partial W_l^{(h)}\over \partial\epsilon}\,{\partial W_{k-l}^{(g-h)}\over \partial\epsilon} \Big)
\eea
If we normalize the moduli as:
\be
t^I =(2\ri \pi)^{1\over 2} \epsilon^I. 
\ee
that last equation reads:
\bea
{\partial W_k^{(g)} \over \partial \overline{t}} &=& 
-\, \kappa {\partial^3 {\overline{F}}^{(0)} \over \partial {\overline{t}}^3} \kappa \,\, 
 {1\over 2}\,\Big( {\partial^2 W_k^{(g-1)}\over \partial t^2} + {\partial \tau\over \partial t} \kappa {\partial W_k^{(g-1)}\over \partial t}  \cr
& & + \,\sum_h \sum_{L\subset K}  {\partial W_l^{(h)}\over \partial t}\,{\partial W_{k-l}^{(g-h)}\over \partial t} \Big)
\eea
For the free energy, we set $k=0$:
\bea
{\partial F^{(g)} \over \partial \overline{t}} &=& 
-\, \kappa {\partial^3 {\overline{F}}^{(0)} \over \partial {\overline{t}}^3} \kappa \,\, 
 {1\over 2}\,\Big( {\partial^2 F^{(g-1)}\over \partial t^2} + {\partial \tau\over \partial t} \kappa {\partial F^{(g-1)}\over \partial t}  + \,\sum_{h=1}^{g-1}  {\partial F^{(h)}\over \partial t}\,{\partial F^{(g-h)}\over \partial t} \Big)
\eea

Since
\be
-\kappa^{IM} {\partial^3 {\overline{F}}^{(0)} \over \partial {\overline{t}}^{\overline M}\partial {\overline{t}}^{\overline N} \partial {\overline{t}}^{\overline K}  } \kappa^{JN}=C^{IJ}_{\overline K}, \qquad {\partial \tau_{JM}\over \partial t^I} \kappa^{KM}=-\Gamma^K_{IJ}
\ee
we recover the holomorphic anomaly equation (\ref{bcov}) for the free energies. The holomorphic 
anomaly equation for the correlation functions can be written  (with the convention that $D_I W_1^{(0)}= du_I$) as
\be
\label{openstring}
\partial_{\overline K} W_k^{(g)} =
 {1\over 2}C_{\overline K}^{IJ} \Big( D_I D_J W_k^{(g-1)}+ \sum_h \sum_{L\subset K}  D_I W_l^{(h)} D_J W_{k-l}^{(g-h)} \Big).
\ee
In the context of topological string theory, these correlation functions correspond to open string amplitudes (see \cite{mm} for 
explicit results in the case of mirrors of toric CY manifolds), therefore the above equations can be regarded as 
holomorphic anomaly equations for open string amplitudes. These equations have been discussed in a general 
context in \cite{bcov, ant}, and they are obtained by looking at the degenerations of Riemann surfaces with boundaries. The result (\ref{openstring}) 
seems to indicates that the only degenerations that contribute in this case are the ones which involve closed string intermediate states. The first term in the r.h.s. 
of (\ref{openstring}) 
corresponds to a node, while the sum in the second term corresponds to all the possible splittings of  
the Riemann surface $\Sigma_{g, k}$ into two Riemann surfaces $\Sigma_{h,l}$ and $\Sigma_{g-h, k-l}$.  

It is also easy to show that the correlation functions (\ref{corrcov}) are given by 
\be
C^{(g)}_{I_1 \cdots I_k}={(-1)^k \over (2\ri \pi)^{k \over 2}} \oint_{\bcycle_{I_1}}  \cdots \oint_{\bcycle_{I_k}} \, W^{(g)}_{k}(p_K)
\ee
since the covariant derivative $ D_{I} = \partial_{I} - \Gamma_I$ commutes with $\oint_\CB$ when acting on
a differential $f(p) \rd p$ with vanishing $\CA$-cycle integrals:
\bea
D_I \oint_{\CB} f(p) \rd p &=& \oint_{\CB} D_I f(p) \rd p + D_I \left(\oint_{\CB} \right) f(p) \rd p \cr
&=& \oint_{\CB} D_I f(p) \rd p -\Gamma_I \oint_{\CB} f(p) \rd p - \partial_I \tau \; \oint_{\underline{\underline\CA}} f(p) \rd p \cr
&=& \oint_{\CB} D_I f(p) \rd p -\Gamma_I \oint_{\CB} f(p) \rd p - \partial_I \tau \,\kappa \; \oint_{\underline{\CB}} f(p) \rd p \cr
&=& \oint_{\CB} D_I f(p) \rd p
\eea
Note that the last equality requires explicitly the condition 
\be
\oint_{\CA} f(p) \rd p = \oint_{\underline{\CA}} f(p) \rd p - \kappa \oint_{\CB} f(p) \rd p =0.
\ee
If we now integrate (\ref{openstring}) over the cycles $B_{I_1}, \cdots, B_{I_k}$, one finds the holomorphic anomaly equation for the correlation function (\ref{holocorr}) in the local case (i.e. without the last line). This indicates again that the open string anomaly equations (\ref{openstring}) involve only the closed string sector in the 
intermediate states.

\subsection{A combinatorial proof}

In the previous section we have given a simple proof of the holomorphic anomaly equations. In \cite{abk} (see also \cite{dewit}) 
it was noticed that these equations have a 
canonical solution of the form 
\be
F^{(g)}(t, \bar t) =F^{(g)} (t) + \Gamma_g(({\rm Im}\, \tau)^{-1}, \partial_{I_1} \cdots \partial_{I_m} F_{r<g}),
\ee
where $\Gamma_g$ is a polynomial functional in $({\rm Im}\, \tau)^{-1}$ and in the derivatives of lower genera holomorphic free energies. 
In \cite{abk}, the coefficients of $\Gamma_g$ were interpreted diagrammatically as the possible degeneracies of a genus $g$ Riemann surface.
In this canonical solution, the antiholomorphic dependence of $F^{(g)}(t, \bar t)$ enters only through ${\rm Im}\, \tau$. We will now show that this  canonical solution 
can be obtained by iterating (\ref{variatW}). This can be used to obtain a second proof of the holomorphic anomaly equations.

\begin{figure}[!ht]
\leavevmode
\begin{center}
\epsfxsize=15cm
\epsfysize=3.2cm
\epsfbox{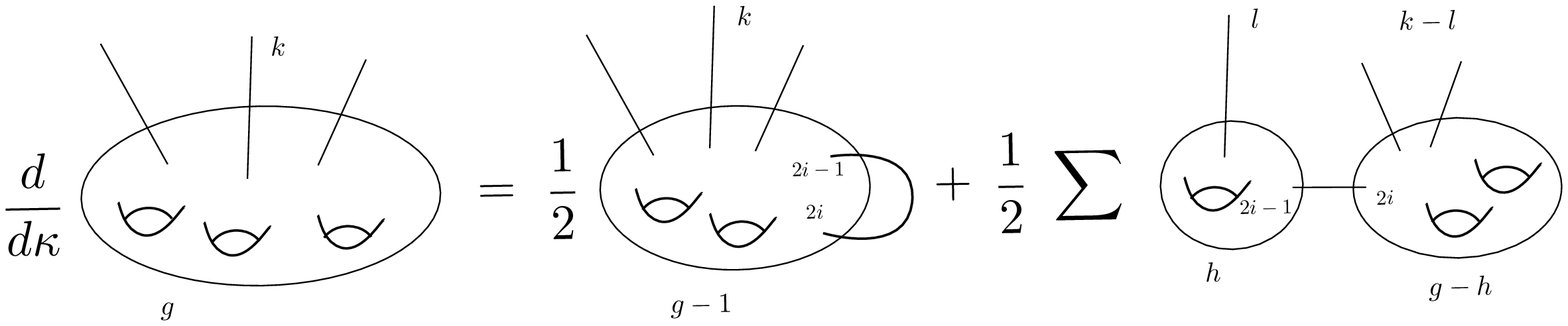}
\end{center}
\caption{A graphic representation of the equation (\ref{variatW}).}
\label{recurrence}
\end{figure}

\begin{figure}[!ht]
\leavevmode
\begin{center}
\epsfxsize=11cm
\epsfysize=12cm
\epsfbox{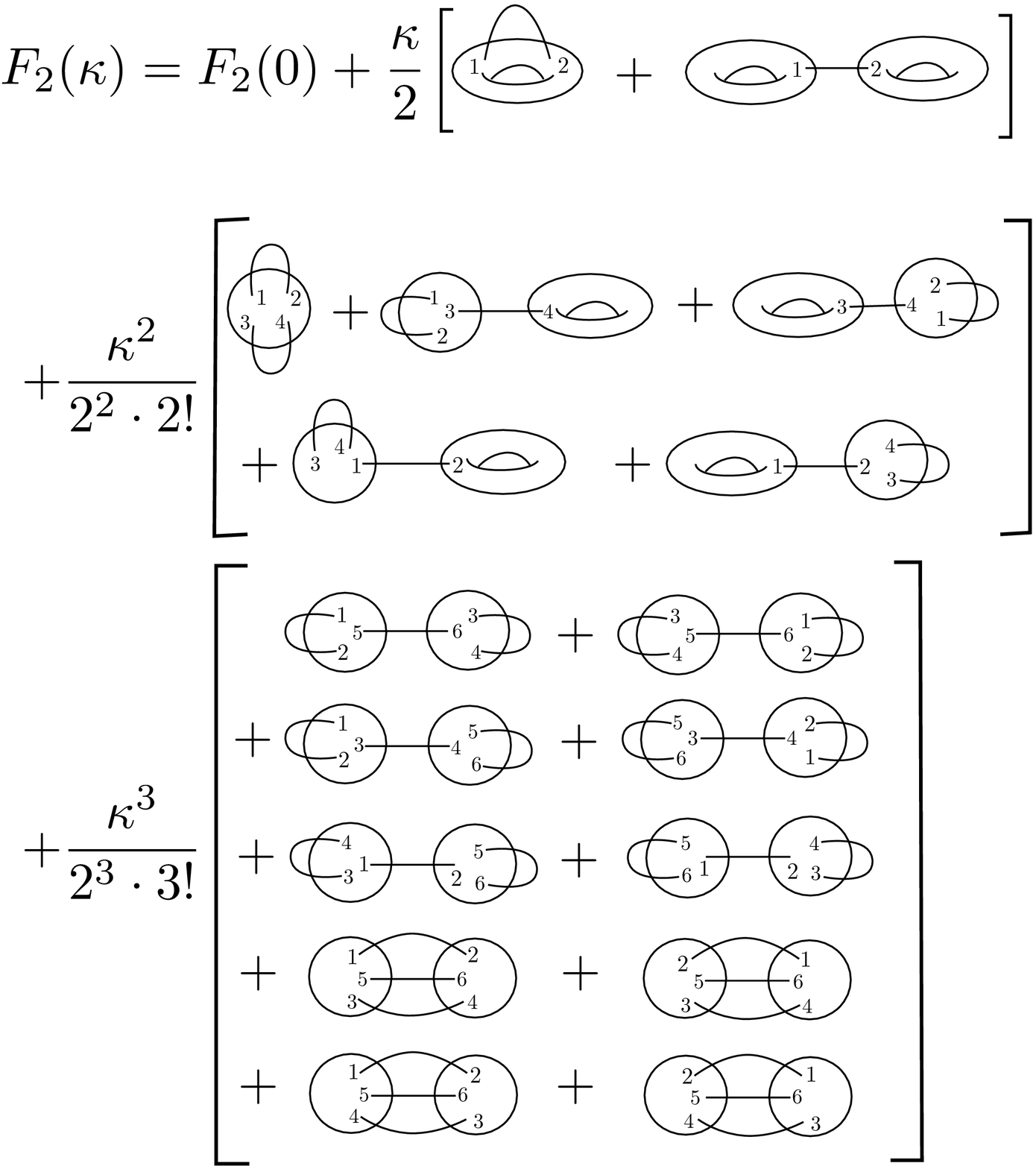}
\end{center}
\caption{The graphs that contribute to $F_2(t, \kappa)$ after iterating \figref{recurrence}.}
\label{genustwo}
\end{figure}

Let $F^{(g)}(t,\kappa=0)$ be the holomorphic free energy of the matrix model at genus $g$. When we turn on $\kappa$ we get the $\kappa$-dependent 
function 
\be
F^{(g)}(t, \kappa)=\sum_{m= 0}^{3g-3} \kappa^m {1\over m!} {\rd^m  \over \rd \kappa^m } F^{(g)}(t, 0).
\ee
The successive derivatives of $F^{(g)}(t, \kappa)$ at $\kappa=0$ can be easily computed by iterating (\ref{variatW}). We take into account 
that at $\kappa=0$ we can substitute the contour integrals of the correlation functions $W_k^g$ by ordinary ({\it not} covariant) derivatives of $F^{(g)}$, as it follows 
from (\ref{seconder}). It is convenient to 
represent (\ref{variatW}) graphically, so that the terms obtained in the iteration are represented by graphs. This representation is shown in \figref{recurrence}. Since the number of legs 
generated in this way is always even, we have labeled the connecting legs by $2i-1$ and $2i$. The result of the iteration is simply (this is easily seen from the graphical representation in \figref{recurrence}):
\be
\label{fgk}
F^{(g)}(t,\kappa)=\sum_{m=0}^{3g-3} \sum_{I_1,\dots,I_{2m}=1}^n \kappa_{I_1,I_2} \dots \kappa_{I_{2m-1},I_{2m}} {1\over 2^m\, m!} \sum_{G_m} A_{G_m},
\ee
where $G_m$ is a connected degenerate surface with $m$ propagators connecting $2m$ points labelled by $1,\dots,2m$, and in such a way that the point labeled by $2i-1$ is connected by a propagator to the point  labeled by $2i$, $i=1, \cdots, m$. 
Each of these surfaces leads to a term $A_{G_m}$. 
These terms are constructed as follows: 
if $G_m$ is made out of $r$ Riemann surfaces of genera $g_1, \cdots, g_r$ with $n_1, \cdots, n_r$ punctures, respectively, (such that $\sum n_i=2m$), and if the $n_i$ punctures of the $i^{\rm th}$ Riemann surface are labeled $J_i=\{ J_{i,1},\dots,J_{i,n_i}\}$, 
then 
\be
\label{agn}
A_{G_m}=\prod_{i=1}^r  \partial_{I_{J_{i,1}}} \cdots  \partial_{I_{J_{i,n_i}}} F^{(g_i)}(t,0).
\ee
As an example of this procedure, we show in \figref{genustwo} the graphs that contribute to $F_2(t, \kappa)$.

\medskip
{\bf Combinatorics}

Notice because of the sum over indices $I_1,\dots, I_{2m}$ in (\ref{fgk}), many graphs $G_m$'s give the same contribution to $F^{(g)}(t,\kappa)$, and thus they better be computed only once with a multiplicity factor.
The multiplicity factor is thus ${\cal N}_{G_m}/(2^m\,m!)$, where ${\cal N}_{G_m}$ is the number of inequivalent ways of relabeling the $2m$ punctures so that the puncture $2i-1$ is linked by a propagator to $2i$ for all $i=1,\dots,m$.
Then notice that:
\beq
{1\over 2^m\, m!} = {(2m-1)!!\over 2m!}
\eeq
and notice that $(2m-1)!!\, {\cal N}_{G_m}$ is the number of inequivalent ways of relabeling the $2m$ punctures (without any constraint on the labels of punctures which are linked together), i.e. it is the number of inequivalent pairings (compatible with $G_m$) of $2m$ points.
This is exactly the multiplicity factor arising from Wick's theorem, i.e. $F^{(g)}(t,\kappa)$ is exactly the Feynman graph expansion of an integral whose propagator is $N^{-2}\kappa$ and whose vertices are the 
$N^{2-2g_i}\,\partial_{I_{J_{i,1}}} \cdots  \partial_{I_{J_{i,n_i}}} F^{(g_i)}(t,0)$.

\smallskip
Notice that because we have chosen $W_1^{(0)}=0$ and $\oint_{\bcycle}\oint_{\bcycle} W_2^{(0)}=0$, $A_m$ vanishes if $G_m$ contains surfaces of genus zero with less than 3 punctures.
Therefore there is no vertices corresponding to the first and second derivative of $F^{(0)}$.

\smallskip
Notice also that the recursive application of \figref{recurrence} generates only connected graphs $G_m$'s, and thus we are in fact computing the Feynman graph expansion of the log of an integral.

\smallskip
Therefore 
\be
F(t, \kappa) = \sum_{g=0}^{\infty} N^{2-2g} F^{(g)}(t, \kappa)
= \ln{(Z(t,\kappa))}
\ee
is exactly the Feynman graph expansion the following integral
\be
\label{genfunanti}
\ba
Z(t,\kappa)
=&  \int \rd \eta
\exp\Bigl\{ -{1\over 2} N^2 (\eta-t) \kappa^{-1} (\eta-t)+ F(\eta,0)- N^2 (\eta -t)^I \partial_I F^{(0)}(t,0 )\\ & \,\,\,\,\, -
{1\over 2} N^2 (\eta-t)^I \partial^2 _{IJ}  F^{(0)}(t,0) (\eta-t)^J \Bigr\}.
\ea
\ee
Indeed, the Feynman graph expansion, is nothing but the saddle point approximation, and here the saddle point equation to large $N$ leading order is 
$\kappa^{-1}_{IJ} (\eta-t)^J -\partial_I F^{(0)}(\eta,0) + \partial_I F^{(0)}(t,0) + \partial^2_{IJ} F^{(0)}(t,0) (\eta-t)^J=0$,
which is clearly satisfied at $\eta=t$.
The propagator is simply 
\be
 {\kappa_{IJ} \over N^2}
\ee
and the vertices are given by 
\be
{N^{2-2g} \over k!} \partial_{I_1}\cdots  \partial_{I_k} F^{(g)} (t),
\ee
where $k\ge 3$ for $g=0$ and $k\ge 1$ for $g\ge 1$.

\bigskip

We will now show that, for the choice $\kappa=(\overline\tau-\tau)^{-1}$, the generating functional (\ref{genfunanti}) satisfies the holomorphic anomaly equations 
of \cite{bcov} in the local case. We first note that there is a cancellation,
\be
\label{simpleint}
Z(t, (\overline\tau-\tau)^{-1}) =  \int \rd \eta
\exp\Bigl\{ -{1\over 2} N^2 (\eta-t) \bar \tau (\eta-t)+ F(\eta,0)- N^2 (\eta -t)^I \partial_I F^{(0)}(t,0 )\Bigr\}
\ee
We compute now: 
\be
{1\over Z} \partial_I Z= N^2\,\,(\bar \tau_{I J}-\tau_{IJ})\, \langle  (\eta-t)^J \rangle + N^2 \partial_I F^{(0)},
\ee
where $\langle \, \rangle$ denotes an average in the integral (\ref{simpleint}). 
It follows that, 
\be
{1\over \hat Z} \partial_I \hat Z=N^2\,\,(\bar \tau_{I J}-\tau_{IJ})\, \langle  (\eta-t)^J \rangle ,
\ee
and
\be
\label{doubled}
{1\over \hat Z} \partial^2_{J I} \hat Z =-N^2 (\bar \tau_{I J}-\tau_{IJ}) -N^2 \partial_J \tau_{IK} \langle (\eta-t)^K \rangle + N^4 (\tau-\bar \tau)_{I K} (\tau-\bar \tau)_{JL} 
\langle (\eta-t)^K (\eta-t)^L\rangle. 
\ee

On the other hand, we compute:
\be
\label{antihol}
{1\over \tilde Z} \partial_{\overline K} \tilde Z= -{1\over 2} N^2 C_{\overline I \overline J \overline K} \langle (\eta-t)^I (\eta -t)^J \rangle +{1\over 2} \partial_{\overline K}  \Delta 
\ee
where 
\be
C_{\overline I \overline J \overline K} = \partial_{\overline K} \bar \tau_{IJ}.
\ee
and
\be
\Delta=\log\, {\rm det}\, \kappa^{-1}
\ee
and we used that the non-holomorphic $F^{(1)}(\kappa)$ generated by the integral (\ref{simpleint}) is
\be
F^{(1)}(t, \kappa) =F^{(1)} (t) + \Delta(\kappa). 
\ee
We express now the quadratic average in (\ref{antihol}) in term of the double derivative (\ref{doubled}). If we use
\be
\partial_{\overline K} \Delta =  \kappa^{IJ} \partial_{\overline K}\bar \tau_{IJ},
\ee
we finally find that
\be
{N^2 \over \tilde Z} \partial_{\overline K} \tilde Z = {1\over 2  \hat Z } C_{\overline K}^{IJ}\Bigl(\partial^2_{IJ} \hat Z -\Gamma^L_{IJ} \partial_L \hat Z\Bigr), 
\ee
where the tensor $C_{\overline K}^{IJ}$ and the Christoffel symbol $\Gamma^L_{IJ} $ are given by  (\ref{tensorlocal}) and (\ref{chris}), respectively. 
In other words we obtain the holomorphic anomaly equations (\ref{bcovbis}) which is equivalent to (\ref{bcov}).

\smallskip

The expansion in \cite{abk} is precisely the one that is obtained by 
expanding the integral (\ref{genfunanti}). This means that the iteration of \figref{recurrence} leads to a representation of $F^{(g)}(t,\kappa)$ which agrees with the canonical 
solution of \cite{abk} for the choice (\ref{choicekappa}).

\section*{Acknowledgments}
We would like to thank Mina Aganagic, Robert Dijkgraaf, Amir Kashani-Poor, Albrecht Klemm, Andrei Okounkov and 
Pierre Vanhove for useful and fruitful discussions on this subject.
This work is partly supported by the Enigma European network MRT-CT-2004-5652, by the ANR project G\'eom\'etrie et int\'egrabilit\'e en physique math\'ematique ANR-05-BLAN-0029-01, by the Enrage European network MRTN-CT-2004-005616, 
by the European Science Foundation through the Misgam program,
by the French and Japaneese governments through PAI Sakurav, by the Quebec government with the FQRNT.


\begin{thebibliography}{99}
\bibliographystyle{plain}


\bibitem{abk}
 M.~Aganagic, V.~Bouchard and A.~Klemm, 
``Topological strings and (almost) modular forms,''
  arXiv:hep-th/0607100.
  %%CITATION = HEP-TH 0607100;%%

\bibitem{Akemann}
G.~Akemann,
``Higher genus correlators for the Hermitian matrix model with multiple 
cuts,''
Nucl.\ Phys.\ B {\bf 482}, 403 (1996)
[arXiv:hep-th/9606004].
%%CITATION = HEP-TH 9606004;%%

\bibitem{ACKM}
J.~Ambjorn, L.~Chekhov, C.~F.~Kristjansen and Y.~Makeenko, 
``Matrix model calculations beyond the spherical limit,''
Nucl.\ Phys.\ B {\bf 404}, 127 (1993)
[Erratum-ibid.\ B {\bf 449}, 681 (1995)]
[arXiv:hep-th/9302014].
%%CITATION = HEP-TH 9302014;%%

\bibitem{ant}
I.~Antoniadis, K.~S.~Narain and T.~R.~Taylor, ``Open string topological amplitudes and gaugino masses,''
  Nucl.\ Phys.\  B {\bf 729}, 235 (2005)
  [arXiv:hep-th/0507244].
  %%CITATION = NUPHA,B729,235;%%

\bibitem{bcov1}
  M.~Bershadsky, S.~Cecotti, H.~Ooguri and C.~Vafa,
  ``Holomorphic anomalies in topological field theories,''
  Nucl.\ Phys.\ B {\bf 405}, 279 (1993)
  [arXiv:hep-th/9302103].
  %%CITATION = HEP-TH 9302103;%%


\bibitem{bcov}
M.~Bershadsky, S.~Cecotti, H.~Ooguri and C.~Vafa, ``Kodaira-Spencer theory of gravity and exact results for quantum string
amplitudes,''
  Commun.\ Math.\ Phys.\  {\bf 165}, 311 (1994)
  [arXiv:hep-th/9309140].
  %%CITATION = HEP-TH 9309140;%%

\bibitem{BertoldiHollowood}
G.~Bertoldi and T.~J.~Hollowood,
  ``Large N gauge theories and topological cigars,''
  JHEP {\bf 0704}, 078 (2007)
  [arXiv:hep-th/0611016].
  %%CITATION = JHEPA,0704,078;%%


\bibitem{BIPZ}
E.~Br\'ezin, C.~Itzykson, G.~Parisi and J.~B.~Zuber,
``Planar Diagrams,''
Commun.\ Math.\ Phys.\  {\bf 59}, 35 (1978).
%%CITATION = CMPHA,59,35;%%

  
\bibitem{CEOloopeq}
 L.~Chekhov, B.~Eynard and N.~Orantin,  ``Free energy topological expansion for the 2-matrix model,''
  JHEP {\bf 0612}, 053 (2006)
  [arXiv:math-ph/0603003].
  %%CITATION = MATH-PH 0603003;%%
  
\bibitem{dewit}
B.~de Wit, G.~Lopes Cardoso, D.~Lust, T.~Mohaupt and S.~J.~Rey, ``Higher-order gravitational couplings and modular forms in N = 2,  D = 4 
heterotic string compactifications,''
  Nucl.\ Phys.\ B {\bf 481}, 353 (1996)
  [arXiv:hep-th/9607184].
  %%CITATION = HEP-TH 9607184;%%

\bibitem{dv}
R.~Dijkgraaf and C.~Vafa,
``Matrix models, topological strings, and supersymmetric gauge theories,''
Nucl.\ Phys.\ B {\bf 644}, 3 (2002)
[arXiv:hep-th/0206255].
%%CITATION = HEP-TH 0206255;%%

\bibitem{dst}
 R.~Dijkgraaf, A.~Sinkovics and M.~Temurhan, ``Matrix models and gravitational corrections,''
  Adv.\ Theor.\ Math.\ Phys.\  {\bf 7}, 1155 (2004)
  [arXiv:hep-th/0211241].
  %%CITATION = HEP-TH 0211241;%%
  
\bibitem{eynloopeq}
 B.~Eynard,
``Topological expansion for the 1-hermitian matrix model correlation
functions,''
arXiv:hep-th/0407261.
%%CITATION = HEP-TH 0407261;%%
 
 \bibitem{eynform}
 B.~Eynard, ``Formal matrix integrals and combinatorics of maps,''
  arXiv:math-ph/0611087.
  %%CITATION = MATH-PH 0611087;%%
 
 \bibitem{EOloopeq}
 B.~Eynard and N.~Orantin, ``Topological expansion of the 2-matrix model correlation functions: Diagrammatic rules for a residue formula,''
  JHEP {\bf 0512}, 034 (2005)
  [arXiv:math-ph/0504058].
  %%CITATION = MATH-PH 0504058;%%
  
\bibitem{eo}
B. Eynard and N. Orantin, ``Invariants of algebraic curves and topological expansion'',
arXiv:math-ph/0702045.

\bibitem{gkmw}
T. Grimm, A. Klemm, M. Mari\~no and M. Weiss, ``Direct integration of the topological string," to appear. 

\bibitem{hk}
 M.~x.~Huang and A.~Klemm, ``Holomorphic anomaly in gauge theories and matrix models,''
  arXiv:hep-th/0605195.
  %%CITATION = HEP-TH 0605195;%%
  
\bibitem{hkq}
M.~x.~Huang, A.~Klemm and S.~Quackenbush, ``Topological string theory on compact Calabi-Yau: Modularity and boundary 
conditions,''
  arXiv:hep-th/0612125.
  %%CITATION = HEP-TH 0612125;%%
  
\bibitem{KazIsing}
V.~A.~Kazakov, ``Ising model on a dynamical planar random lattice: Exact solution,''
  Phys.\ Lett.\ A {\bf 119}, 140 (1986).
  %%CITATION = PHLTA,A119,140;%%
\bibitem{selfdual}
 A.~Klemm, W.~Lerche, P.~Mayr, C.~Vafa and N.~P.~Warner, 
 ``Self-Dual Strings and N=2 Supersymmetric Field Theory,''
  Nucl.\ Phys.\ B {\bf 477}, 746 (1996)
  [arXiv:hep-th/9604034].
  %%CITATION = HEP-TH 9604034;%%

\bibitem{kmt}
A.~Klemm, M.~Mari\~no and S.~Theisen, ``Gravitational corrections in supersymmetric gauge theory and matrix models,''
  JHEP {\bf 0303}, 051 (2003)
  [arXiv:hep-th/0211216].
  %%CITATION = HEP-TH 0211216;%%
  
\bibitem{kz}
  A.~Klemm and E.~Zaslow, ``Local mirror symmetry at higher genus,''
  arXiv:hep-th/9906046.
  %%CITATION = HEP-TH 9906046;%%
  
 \bibitem{mmhouches}
  M.~Mari\~no, ``Les Houches lectures on matrix models and topological strings,''
  arXiv:hep-th/0410165.
  %%CITATION = HEP-TH 0410165;%%

\bibitem{mm}
M.~Mari\~no, ``Open string amplitudes and large order behavior in topological string theory,''
  arXiv:hep-th/0612127.
  %%CITATION = HEP-TH 0612127;%%

 \bibitem{nv}
 A.~Neitzke and C.~Vafa, 
 ``Topological strings and their physical applications,''
  arXiv:hep-th/0410178.
  %%CITATION = HEP-TH 0410178;%%

\bibitem{nekrasov}
N.~A.~Nekrasov, ``Seiberg-Witten prepotential from instanton counting,''
  Adv.\ Theor.\ Math.\ Phys.\  {\bf 7}, 831 (2004)
  [arXiv:hep-th/0206161].
  %%CITATION = 00203,7,831;%%

\bibitem{thooft}
G.~'t Hooft, ``A planar diagram theory for strong interactions,''
Nucl.\ Phys.\ B {\bf 72} (1974) 461.
%%CITATION = NUPHA,B72,461;%%


\bibitem{vonk}
 M.~Vonk, ``A mini-course on topological strings,''
  arXiv:hep-th/0504147.
  %%CITATION = HEP-TH 0504147;%%
  

\bibitem{witten}
E.~Witten, ``Quantum background independence in string theory,'' \\
  arXiv:hep-th/9306122.
  %%CITATION = HEP-TH/9306122;%%








\end{thebibliography}
\end{document}